\documentclass[12pt]{article}
\usepackage{amssymb,amsmath,amsthm}
\usepackage{hyperref}
\usepackage{array,longtable}
\usepackage[dvips]{graphicx}
\usepackage[dvips]{color}

\usepackage[mathscr]{eucal}

\renewcommand{\theequation}{\thesection.\arabic{equation}}

\newcommand{\be}{\begin{equation}}
\newcommand{\ee}{\end{equation}}
\newcommand{\bes}{\begin{equation*}}
\newcommand{\ees}{\end{equation*}}

\newcolumntype{V}{>{$}m{4cm}<{$}}
\newcolumntype{C}{>{$}c<{$}}
\newcolumntype{L}{>{$}l<{$}}
\newcolumntype{R}{>{$}r<{$}}

\newcommand{\Zh}{\mathbb Z}

\newcommand{\ra}{\rightarrow}

\newcommand{\Qc}{\mathcal{Q}}

\newcommand{\pd}{\partial}

\newcommand{\la}{\langle\!\langle}
\renewcommand{\ra}{\rangle\!\rangle}

\begin{document}

\thispagestyle{empty}

\vspace*{19.0ex}

\centerline{\huge \textsc{NS Ghost Slivers}}


\vspace*{8.0ex}

\centerline{\large \rm I.Ya.~Aref'eva$^{\dag}$,
 A.A.~Giryavets$^{\ddag}$ and A.S.~Koshelev$^{\dag}$}

\vspace*{7.0ex}
\centerline{\large \it ~$^{\dag}$Steklov Mathematical Institute,
 Russian Academy of Sciences,}
\centerline{\large \it Gubkin st. 8, Moscow, Russia, 117966 }
\centerline{\texttt{arefeva@mi.ras.ru, koshel@orc.ru}}
\vspace*{2.5ex}
\centerline{\large \it $^{\ddag}$Faculty of Physics, Moscow State
University,} \centerline{\large \it Moscow, Russia, 119899 }
\centerline{ \texttt{alexgir@mail.ru}}
\vspace*{2.5ex}
\vspace*{9.0ex}
\centerline{\bf Abstract}
\medskip

Neveu-Schwarz ghost slivers in pictures zero and minus one  are constructed.
In particular, using algebraic methods $\beta$, $\gamma$
ghost sliver in the $-1$ picture is obtained.
The algebraic method consists  in solving a projector equation
in an algebra, where the multiplication is defined by a pure 3-string vertex
without any insertions at the string midpoint.
We show that this projector is a sliver in a twisted version of
$\beta$, $\gamma$ conformal theory.
We also show that the product of the twisted $b$, $c$  and
$\beta$, $\gamma$ ghost slivers solves an equation
that appears after a  special rescaling  of  super VSFT.
\vfill \eject

\baselineskip=16pt

\newpage
\tableofcontents

\section{Introduction.}
\label{sec:intro}
\setcounter{equation}{0}

 During the last year the bosonic vacuum string field theory (VSFT)
 proposed to describe physics around  the bosonic tachyon vacuum
  \cite{0012251} has been  investigated in many papers
\cite{RZ}-\cite{0201159}.
By product this study has revealed many interesting features
of string field algebra.
The characteristic feature of VSFT is a very simple form of the
kinetic operator $\Qc$. This $\Qc$ is just $c(i)-c(-i)$
in the bosonic case.
 It is apparently related
 by a singular transformation with the shifted BRST operator,
 $Q=Q_B-[A_0,\cdot]$, where $A_0$ is a vacuum solution of
 the open string field theory (SFT) describing the D-brane decay.
Due to an absence of a dependence of $\Qc$ on the matter fields
VSFT equations of motion
admit a  factorized form with the projector-like matter part.
Solutions of projector equations have been discussed in many details
\cite{RZ}-\cite{0201197}.
These equations are similar to the non-commutative soliton
equations  in the large non-commutativity limit \cite{GMS}.

Ghost part of VSFT equations of motion has been studied in
\cite{0108150,0110124,0111034,0111087,0111129,0201015,0201177}.
 It was observed that a sliver constructed in the twisted conformal theory
with new $SL(2,\mathbb{R})$ invariant vacuum
solves the ghost part of VSFT  equation of motion.
This equation is a usual SFT  equation of motion with a canonical choice
of ghost kinetic term
that is a local insertion at the  string midpoint.

It is interesting to consider a generalization of these treatments
to the case of superstrings. A generalization of VSFT
to superstrings has been discussed in \cite{0012251} and more recently in
\cite{0112214}
and \cite{0112231} in the context of cubic SSFT \cite{AMZ1,PTY} and
non-polynomial SSFT \cite{0002211},
respectively. Fermionic projectors, such as the NS sliver,
have been constructed   in \cite{0112214,0112231}.
Although in the matter sector we  have standard
equations for projectors, in the ghost sector we have to solve a bit more complicated equation.
As it has been noted in \cite{0201197},
the VSFT
kinetic operator $\hat{\Qc }$ in superstring case
inevitably has  the matrix structure,
or in component notations has a non-diagonal form
and
mixes GSO$\pm$ sectors:
\begin{equation}
\hat{\Qc}=\begin{pmatrix}
  \Qc_{\textsf{odd}} &\Qc_{\textsf{even}} \\
  -\Qc_{\textsf{even}} & -\Qc_{\textsf{odd}}
  \end{pmatrix},
  \label{QQcal}
  \end{equation}
with $\Qc_{\textsf{even}}\neq 0 $.
This is related with the fact that the corresponding $A_0$
describing a decay of a non-BPS brane has   non-zero  GSO$-$ component.
If $\Qc_{\textsf{even}}$  were zero,
we could take $\Qc_{\textsf{odd}}$ to be the ghost kinetic
operator used in the bosonic VSFT \cite{0111129}.
 In \cite {0201197}  the following candidates for $\Qc_{\textsf{odd}}$
   and  $\Qc_{\textsf{even}}$  have  been  proposed:
   \begin{equation}
\Qc_{\textsf{odd}}=c(i)+\frac{1}{2\pi i}\oint b(z)\gamma^2(z)dz,~~~~~
\Qc_{\textsf{even}}=\gamma(i).
\label{formal-q-int}
\end{equation}

Therefore, VSFT equations of motion in supersting case are more complicated
in comparison with their bosonic  analog. Under an assumption of a special
splitting of these equations one however can get projector-like
equations with a ghost insertion. To solve these equations it is reasonable
to consider slivers in different pictures.
A presence of different pictures (Bose seas) is a special feature of the
$\beta$, $\gamma$ system.

A study of superghost slivers and wedge states
is a subject of the present paper. We perform this study using
generalizations of different technics known in bosonic case.
Sometimes these generalizations are straightforward, but sometimes they need
new  calculations.Since the VSFT in the ghost sector
deals with the projector-like  equation with an insertion it is
appropriate to study slivers in  different pictures.

We start from an algebraic method. The algebraic method consists in
solving a projector equation
in a $*^{\prime}$-algebra,
\begin{equation}
|\Xi^{\beta\gamma}\rangle *{}'|\Xi^{\beta\gamma}\rangle=
|\Xi^{\beta\gamma}\rangle.\label{intr?}
\end{equation}
 where
the multiplication $*^{\prime}$ is defined by a pure 3-string vertex  without
any insertion, $*$ is reserved for a standard multiplication
with a suitable ghost insertion  \cite{Witten,AMZ1,PTY,0111208}.
Then we construct a twisted CFT for $\beta$, $\gamma$
ghosts ("CFT in $-1$ picture") and  show that twisted
$\beta$, $\gamma$ sliver  solves
the projector-like equation (\ref{intr?}).

We will  argue that
the direct product of twisted $b,c$ and twisted
$\beta$, $\gamma$ slivers
\begin{gather}
|\Xi\rangle=|\Xi'{}_{bc}\rangle\otimes|\Xi'{}_{\beta\gamma}\rangle,
\label{xi-xi-xi}
\end{gather}
solves equation
\begin{gather}
(c(i)e^{-\phi(i)}-c(-i)e^{-\phi(-i)})|\Xi\rangle+|\Xi\rangle*|\Xi\rangle=0.
\label{int-sliv-minus1}
\end{gather}
Equation \eqref{int-sliv-minus1} defines  the ghost
sliver in the $-$1 picture  and
 appears after a  special rescaling of super VSFT(VSSFT).

The paper is organized as follows.
In section 2 we give an algebraic construction of
the fermionic ghost sliver and wedge states in the minus one
picture.
In Section 3  following \cite{0111129} we introduce
the  twisted  $\beta$, $\gamma$
CFT
and find ghost slivers in  pictures -1 and 0.
Further we use the twisted  CFT  to show that the sliver
\eqref{xi-xi-xi} satisfies  equation
\eqref{int-sliv-minus1}.
 In section 4 it is shown that this equation arises as
one of the equations of motion for super VSFT action.

\section{Algebraic Construction.}
\setcounter{equation}{0}
In this section we will construct algebraically the  $\beta, \gamma$ ghost sliver
in the $-1$ picture. It is convenient to start from  this picture  because
in this picture $\beta$,$\gamma$ annihilation and creation
operators are defined symmetrically.
It will be shown  that the matrix of this sliver coincides with the one
of the fermionic matter sliver.

\subsection{Algebraic construction of sliver}
The fermionic ghosts have the following mode expansion
\begin{gather}
\beta_{\pm}(\sigma)=\sum_{r}\beta_{r}e^{\pm ir\sigma},\quad
\gamma_{\pm}(\sigma)=\sum_{r}\gamma_{r}e^{\pm ir\sigma},
\end{gather}
where $r\in \mathbb{Z}+1/2$, $\sigma\in[0,\pi]$.
The modes $\beta_r,\,\gamma_r$ have the following commutation
relations
\begin{equation}
[\gamma_{r},\beta_{s}]=\delta_{r+s,\,0}.\label{comm-bg}
\end{equation}
Ghosts $\beta(\sigma)$ and $\gamma(\sigma)$ have weights
$3/2$ and $-1/2$ correspondingly.
Representation of the commutation relations \eqref{comm-bg} is specified by
a chosen picture.
$|q\rangle$ is
a vacuum in the $q$ picture which is defined as
\begin{subequations}
\begin{gather}
\beta_{s}|q\rangle=0,\quad s>-q-3/2,\\
\gamma_{r}|q\rangle=0,\quad r\geq q+3/2.
\end{gather}
\end{subequations}
The adjoint of the $q$-vacuum is a state $\langle-q-Q|$ such that
$\langle-q-Q|q\rangle=1$ with $Q=2$.
A distinguished feature of the -1 picture is that the adjoint vacuum
has the same $q$-charge. Due to this property in the -1 picture one
can define superghost state multiplication without any additional insertions
\begin{equation}
(|A\rangle*{}'|B\rangle)_{3}= {}_{1}\langle A|_{2}\langle B|V^{\beta\gamma}\rangle_{123},
\end{equation}
where $|V^{\beta\gamma}\rangle$ is the three string vertex in
the minus one picture.
An algebraic construction
of the fermionic ghost vertices: identity, reflector and
three string vertex \cite{GrJe3} over the vacuum in the $-1$ picture
is reviewed in Appendix A.
We will use these  vertices in our construction of the sliver in the $-1$ picture.

As in the NS matter case instead of finding a solution to
projector equation
\begin{equation}
|\Xi^{\beta\gamma}\rangle*{}'|\Xi^{\beta\gamma}\rangle=
|\Xi^{\beta\gamma}\rangle,\label{intr-sliv-eqXIXI-intr}
\end{equation}
we will first construct a solution to the linear equation
\begin{equation}
|\Xi^{\beta\gamma}\rangle*{}'|-1\rangle=|\Xi^{\beta\gamma}\rangle.
\label{intr-sliv-eqXI0}
\end{equation}
As in the matter fermionic case equations
\eqref{intr-sliv-eqXI0} and (\ref{intr-sliv-eqXIXI-intr})
have the same nontrivial solutions.

We look for the   $\beta$,$\gamma$ ghost
sliver in the following squeezed form
\begin{equation}
|\Xi^{\beta\gamma}\rangle=\mathcal{N}^{\beta\gamma}
\exp(\beta_{-r}S^{\beta\gamma}_{rs}\gamma_{-s})|-1\rangle,
\end{equation}
where $\mathcal{N}^{\beta\gamma}$ is a normalization constant and
$|-1\rangle$ is a vacuum in the fermionic ghost sector in the $-1$ picture.
The following formulae for the squeezed states multiplication for
fermionic ghosts holds
\begin{multline}
\langle -1|\exp(-\beta_{r}
S_{rs}\gamma_{s})\exp(\mu_{r}\beta_{-r}+\nu_{r}\gamma_{-r}+
\beta_{-r}V_{rs}\gamma_{-s})|-1\rangle\\
=\det(1-S_{rl}V_{ls})^{-1}\exp(\nu_{r}(1-S_{rl}V_{lk})^{-1}S_{ks}\mu_{s}).\label{sqv-betagamma}
\end{multline}

Using reflector \eqref{refl} one gets
the BPZ conjugated state $\langle\Xi^{\beta\gamma}|$ of $|\Xi^{\beta\gamma}\rangle$
\begin{equation}
\langle\Xi^{\beta\gamma}|=\mathcal{N}^{\beta\gamma}\langle -1|
\exp(-\beta_{r}(CS^{\beta\gamma}C)_{rs}\gamma_{s}).
\end{equation}
Equation (\ref{intr-sliv-eqXI0}) can be rewritten in the form
\begin{equation}
_{1}\langle\Xi^{\beta\gamma}|_{2}\langle
-1|V^{\beta\gamma}\rangle_{123}=|\Xi^{\beta\gamma}\rangle_{3}.
\label{betagamma-1}
\end{equation}
Using \eqref{sqv-betagamma} one gets from
(\ref{betagamma-1})  the following equation
\begin{equation}
M^{\beta\gamma}_{21}(1-T^{\beta\gamma}M^{\beta\gamma}_{11})^{-1}T^{\beta\gamma}M^{\beta\gamma}_{12}+M^{\beta\gamma}_{11}=T^{\beta\gamma}.\label{bg-sl-eq}
\end{equation}
Here we denote $T^{\beta\gamma}=CS^{\beta\gamma}$, where $C_{rs}=(-1)^{r}\delta_{rs}$.  Assuming the
following commutation relations
\begin{equation}
[T^{\beta\gamma},M^{\beta\gamma}_{ab}]=0,\quad \forall\;\; a,b,
\end{equation}
and using the properties \eqref{matrix-gh-prop}, equation
\eqref{bg-sl-eq} can be rewritten as
\begin{equation}
T^{\beta\gamma\,2}M^{\beta\gamma}_{11}-T^{\beta\gamma}(1+CI^{\beta\gamma\,-1}M^{\beta\gamma}_{11})+M^{\beta\gamma}_{11}=0.
\end{equation}
Explicit solutions to this equation are
\begin{equation}
T^{\beta\gamma}_{\pm}=\frac{1+CI^{\beta\gamma\,-1}M^{\beta\gamma}_{11}\mp\sqrt{(1+CI^{\beta\gamma\,-1}M^{\beta\gamma}_{11})^{2}-4M^{\beta\gamma\,2}_{11}}}{2M^{\beta\gamma}_{11}}.
\label{sliv-gh}
\end{equation}
The projector equation
$$_{1}\langle\Xi^{\beta\gamma}|_{2}\langle\Xi^{\beta\gamma}
|V^{\beta\gamma}\rangle_{123}=|\Xi^{\beta\gamma}\rangle_{3}$$
together with  \eqref{sqv-betagamma}  gives
the following equation
\begin{equation}
(M^{\beta\gamma}_{12},M^{\beta\gamma}_{21})\left(1-T^{\beta\gamma}\begin{pmatrix}
  M^{\beta\gamma}_{11} & M^{\beta\gamma}_{12} \\
  M^{\beta\gamma}_{21} & M^{\beta\gamma}_{22}
\end{pmatrix} \right)^{-1}\begin{pmatrix}
  T^{\beta\gamma}M^{\beta\gamma}_{21} \\
  T^{\beta\gamma}M^{\beta\gamma}_{12}
\end{pmatrix}+M^{\beta\gamma}_{11}=T^{\beta\gamma}.
\end{equation}
This equation can be rewritten in the form
\begin{equation}
(T^{\beta\gamma}-CI^{\beta\gamma})(T^{\beta\gamma\,2}M^{\beta\gamma}_{11}-T^{\beta\gamma}(1+CI^{\beta\gamma\,-1}M^{\beta\gamma}_{11})+M^{\beta\gamma}_{11})=0.
\end{equation}
One of the solutions to this equation is the identity and the other ones,
as has been mentioned above, coincide
with solutions of \eqref{sliv-gh}.
Expressions \eqref{sliv-gh} can be rewritten in terms of the matrices $F$,
$\tilde{F}$ (see Appendix A) and $C$ as
\begin{gather}
T^{\beta\gamma}_{\pm}=-\frac{1}{F}(C\tilde{F}\pm i).
\end{gather}
So we obtain that the matrix $S^{\beta\gamma}=-CT^{\beta\gamma}$ is identical
to that of matter fermions. This is not surprising and has an origin
 in the fact that
the fermionic ghost vertices can be obtained
from the fermionic matter vertices by changing the sings of $F$, $\tilde{F}$
and $C$. Moreover the equation for the fermionic ghost sliver is identical to that for
fermionic matter sliver.

\subsection{Algebraic construction of wedge states}
Here we give an algebraic construction of the fermionic ghost wedge states
in the minus one picture following steps of \cite{0107101} for the bosonic case.
The algebra obeyed by the wedge states is
\begin{equation}
|n\rangle*{}'|m\rangle=|n+m-1\rangle.
\end{equation}
We look for the ghost wedge states of the following form
\begin{gather}
|n\rangle=\mathcal{N}_{n}^{\beta\gamma}\exp(-\beta_{-r}(CT^{\beta\gamma}_{n})_{rs}\gamma_{-s})|-1\rangle.
\end{gather}
The recursion for the wedge states exponential factors and norms
can be explicitly written using the relation
\begin{equation}
|n\rangle*{}'|2\rangle=|n+1\rangle,
\end{equation}
where $|2\rangle$ corresponds to the vacuum in the minus one picture.
For the ghost wedges one gets
\begin{gather}
T^{\beta\gamma}_{n+1}=\frac{M^{\beta\gamma}_{11}(1-CI^{\beta\gamma\,-1}T^{\beta\gamma}_{n})}{1-T^{\beta\gamma}_{n}M^{\beta\gamma}_{11}},\\
 \mathcal{N}^{\beta\gamma}_{n+1}=\mathcal{N}^{\beta\gamma}_{n}\det\left(\frac{1-M^{\beta\gamma}_{11}T^{\beta\gamma}_{n}}{1-M^{\beta\gamma}_{11}T^{\beta\gamma}}\right)^{-1}.
\end{gather}
Using
\begin{equation}
M^{\beta\gamma}_{11}=\frac{T^{\beta\gamma}}{1-CI^{\beta\gamma\,-1}T^{\beta\gamma}+T^{\beta\gamma\,2}}=\frac{(CI^{\beta\gamma})^5-CI^{\beta\gamma}}{(1-(CI^{\beta\gamma})^{2})(1-3(CI^{\beta\gamma})^{2})}
\end{equation}
one finds
\begin{gather}
T^{\beta\gamma}_{n}=T^{\beta\gamma}\frac{1-\Upsilon^{n-2}}{T^{\beta\gamma\,2}-\Upsilon^{n-2}},\\
\mathcal{N}^{\beta\gamma}_{n}=\det\left(\Upsilon^{-n+2}\frac{-T^{\beta\gamma}}{T^{\beta\gamma}_{n}-T^{\beta\gamma}}\right)^{-1}
=\det\left(\frac{1-T^{\beta\gamma\,2}\Upsilon^{-n+2}}{1-T^{\beta\gamma\,2}}\right)^{-1},
\end{gather}
where
\begin{equation}
\Upsilon=-\frac{(CI^{\beta\gamma}-T^{\beta\gamma})}{T^{\beta\gamma}(1-CI^{\beta\gamma}T^{\beta\gamma})}.
\end{equation}
For the fermionic matter wedges
\begin{gather}
|n\rangle_{m}=\mathcal{N}_{n}^{10}\exp(-\frac{1}{2}\psi^{\dag}CT_{n}\psi^{\dag})|0\rangle,
\end{gather}
we get the same equations, but the matrices
$F$, $\tilde{F}$, $C$ should be interchanged with $-F$, $-\tilde{F}$, $-C$
and the power in norm should be changed from $-1$ to $1/2$.

\section{Conformal Construction.}
\setcounter{equation}{0}

In this section we present the twisted superghost conformal theory
and derive corresponding equations in analogy with the one
constructed by Gaiotto, Rastelli, Sen and
Zwiebach \cite{0111129}.

\subsection{Twisted CFT}
A twisted CFT is introduced by subtracting from the stress energy tensor $T(w)$
of the $(\beta,\gamma)$ system the derivative of $U(1)$ ghost number current
$j$ as follows
\begin{gather}
T'(w)=T(w)-\pd j(w),\quad \Bar{T}'(\Bar{w})=\Bar{T}(\Bar{w})-\pd\Bar{j}(\Bar{w}),\quad j=-\beta\gamma.
\end{gather}
More explicitly for the holomorphic stress energy tensor one obtains
\begin{eqnarray}
T(w)=-\frac32\beta\pd\gamma(w)-\frac12\pd\beta\gamma(w),\quad \mathrm{with~}c=11,\\
T'(w)=-\frac12\beta'\pd\gamma'(w)+\frac12\pd\beta'\gamma'(w),\quad
\mathrm{with~}c=-1,
\end{eqnarray}
where $(\beta',\gamma')$ denotes the superghosts of the twisted CFT and $c$ is the central charge.
Due to this modification the weights of the $\beta'$ and $\gamma'$ become equal
to $1/2$
\begin{subequations}
\begin{gather}
T'(w)\beta'(z)=\frac{1}{2}\frac{\beta'(w)}{(z-w)^{2}}+\frac{\pd\beta'(w)}{z-w},\\
T'(w)\gamma'(z)=\frac{1}{2}\frac{\gamma'(w)}{(z-w)^{2}}+\frac{\pd\gamma'(w)}{z-w},
\end{gather}
\end{subequations}
and the superghost current $j'=-\beta'\gamma'$ has no anomaly.
Fermionic ghosts in the original theory are bosonised as
\begin{gather}
\gamma(w)=\eta e^{\phi}(w),\quad
\beta(w)= e^{-\phi}\pd\xi(w),
\end{gather}
so that the ghost number current is expressed in the form $j=-\pd\phi$.
The Euclidean world-sheet actions
$S$ and $S'$ for the fields $\phi$ and $\phi'$ correspondingly are related as
\begin{gather}
S[\phi]= S'[\phi]-\frac{1}{2\pi}\int_{\Sigma}
d^{2}\zeta\;\sqrt{g}R^{(2)}(\phi+\Bar{\phi}),
\end{gather}
where $\zeta$ denotes the world-sheet coordinates, $g$ denotes the
Euclidean world-sheet metric and $R^{(2)}$ is the scalar
curvature.

We assume that scalar curvature is proportional to $\delta$-function,
which has a support on the infinity in some coordinates on $\Sigma$.
Therefore we can identify the fields $\phi$ and $\phi'$ of two CFTs.
The states in the two theories can be identified by
the following map between the oscillators and the vacuum states
\begin{gather}
\beta_{n}\leftrightarrow\beta'{}_{n},\quad \gamma_{n}\leftrightarrow\gamma'{}_{n},\quad
|-1\rangle\leftrightarrow|0'\rangle,\quad \langle
-1|\leftrightarrow\langle 0'|,\quad
\langle0'|0'\rangle=1,\label{cooresp}
\end{gather}
where $|0\rangle$ and $|0'\rangle$ are the $SL(2,\mathbb{R})$
invariant vacua  of two  theories and $|-1\rangle=e^{-\phi(0)}|0\rangle$.

In the CFT$'$ the fields $\beta',\gamma'$ are bosonized as in the original
theory
\begin{gather}
\gamma'(w)=\eta e^{\phi}(w),\quad
\beta'(w)= e^{-\phi}\pd\xi(w).
\end{gather}
Notice that we do not introduce new notations for the
$(\xi,\eta)$ system because it has  not changed.

One gets the following operator product expansions
\begin{subequations}
\begin{gather}
T'(z)e^{\pm\phi}(w)=-\frac{1}{2}\frac{e^{\pm\phi}}{(z-w)^{2}}+\frac{\pd e^{\pm\phi}}{z-w},\\
T'(z)e^{\pm\phi/2}(w)=-\frac{1}{8}\frac{e^{\pm\phi/2}}{(z-w)^{2}}+\frac{\pd e^{\pm\phi/2}}{z-w}.
\end{gather}
\end{subequations}

\subsection{Fermionic ghost surface states}
In this subsection we construct superghost surface states using CFT methods.
The advantage of the CFT method in comparison with the operator method,
that we have used in  Section 2, is that we do not have
to postulate the sliver equation from the very beginning.
The aim of this section is to define a sliver state as a surface state
over $SL(2,\mathbb{R})$ invariant vacuum in CFT and  CFT$'$, correspondingly,
by the conformal map used in the matter case.

First we define the surface state for the original $(\beta,\gamma)$ system.
The fermionic ghost surface state
corresponding to the conformal map $\lambda(\xi)$ is defined as
\begin{gather}
\langle\Lambda|=\mathcal{N}_{\beta\gamma} \langle 0|\exp(-\sum_{r\geq 3/2\atop s\geq -1/2}\gamma_{r}\Lambda_{rs}\beta_{s}),
\end{gather}
where $\mathcal{N}_{\beta\gamma}$ is a normalization factor and
the matrix $\Lambda_{rs}$ is defined so that
the following identity holds
\begin{gather}
\langle 0|\exp(-\sum_{r\geq 3/2 \atop s\geq
-1/2}\gamma_{r}\Lambda_{rs}\beta_{s})\gamma(w)
\beta(z)e^{-Q\phi(0)}|0\rangle=
\langle \lambda\circ \gamma(w)\lambda\circ
\beta(z)\lambda\circ e^{-Q\phi(0)}\rangle.
\label{def-state-map-h-2}
\end{gather}

One can evaluate $\Lambda_{rs}$
explicitly. To this end one has to calculate the left hand side and
right hand side of \eqref{def-state-map-h-2}.
Substitution of
$\gamma(w)=\sum_{r}\gamma_{-r}w^{r+1/2}$ and
$\beta(z)=\sum_{s}\beta_{-s}z^{s-3/2}$ into the left hand side of \eqref{def-state-map-h-2}
yields
\begin{gather}
h(z,w)\equiv\langle 0|\exp(-\gamma_{r}\Lambda_{rs}\beta_{s})\gamma(w)\beta(z)e^{-Q\phi(0)}|0\rangle= -\sum_{r,s}w^{r+1/2}z^{s-3/2}\Lambda_{rs},\end{gather}
therefore
\begin{gather}
\Lambda_{rs}=-\oint\frac{dz}{2\pi i}\frac{1}{z^{r-1/2}}
\oint\frac{dw}{2\pi i}\frac{1}{w^{s+3/2}}h(z,w).
\end{gather}

Further one evaluates the correlation function in the right hand side of
\eqref{def-state-map-h-2}
\begin{multline}
\langle \lambda\;\circ\; \gamma(w)\lambda\;\circ\; \beta(z)\lambda\;\circ\; e^{-Q\phi(0)}\rangle\\
=\langle \left(\frac{\pd\lambda(w)}{\pd w}\right)^{-1/2}\gamma(\lambda(w))
 \left(\frac{\pd\lambda(z)}{\pd z}\right)^{3/2}\beta(\lambda(z)) e^{-Q\phi(\lambda(0))}\rangle\\
=\left(\frac{\pd\lambda(w)}{\pd w}\right)^{-1/2}
\left(\frac{\pd\lambda(z)}{\pd z}\right)^{3/2}
\langle \eta e^{\phi}(\lambda(w)) e^{-\phi} \pd\xi(\lambda(z)) e^{-Q\phi(\lambda(0))} \rangle\\
=\left(\frac{\pd\lambda(w)}{\pd w}\right)^{-1/2}
\left(\frac{\pd\lambda(z)}{\pd z}\right)^{3/2}
\frac{1}{\lambda(w)-\lambda(z)}\left(\frac{\lambda(w)-\lambda(0)}{\lambda(z)-\lambda(0)}\right)^{-Q}.
\end{multline}
One gets the following answer
\begin{gather}
\label{lambda-conf-sliver0}
\Lambda_{rs}=\oint\frac{dz}{2\pi i}\frac{1}{z^{r-1/2}}
\oint\frac{dw}{2\pi i}\frac{1}{w^{s+3/2}}
\left(\frac{\pd\lambda(w)}{\pd w}\right)^{-1/2}
\left(\frac{\pd\lambda(z)}{\pd z}\right)^{3/2}
\frac{1}{\lambda(z)-\lambda(w)}
\left(\frac{\lambda(w)-\lambda(0)}{\lambda(z)-\lambda(0)}\right)^{-2}.
\end{gather}

The fermionic ghost surface state in CFT'
corresponding to the conformal map $\lambda(\xi)$ is defined as
\begin{gather}
\langle\Lambda'|=\mathcal{N}'{}_{\beta\gamma} \langle 0'|\exp(-\sum_{r\geq 1/2\atop s\geq 1/2}\gamma_{r}\Lambda'{}_{rs}\beta_{s}),
\end{gather}
where $\mathcal{N}'{}_{\beta\gamma}$ is a normalization factor and
the matrix $\Lambda'{}_{rs}$ is defined so that
the following identity holds
\begin{gather}
\langle 0'|\exp(-\sum_{r\geq 1/2 \atop s\geq
1/2}\gamma_{r}\Lambda'{}_{rs}\beta_{s})\gamma'(w)
\beta'(z)|0'\rangle=
\langle \lambda\circ \gamma'(w)\lambda\circ
\beta'(z)\rangle'.
\label{def-state-map-h-2-1}
\end{gather}

Substitution of
$\gamma'(w)=\sum_{r}\gamma_{-r}w^{r-1/2}$ and
$\beta'(z)=\sum_{s}\beta_{-s}z^{s-1/2}$ into the left hand side of \eqref{def-state-map-h-2-1}
 yields
\begin{gather}
h'(z,w)\equiv\langle 0'|\exp(-\sum_{r\geq 1/2 \atop s\geq
1/2}\gamma_{r}\Lambda'{}_{rs}\beta_{s})\gamma'(w)
\beta'(z)|0'\rangle=-\sum_{r,s}w^{r-1/2}z^{s-1/2}\Lambda'{}_{rs},
\label{hminusone-1}
\end{gather}
therefore
\begin{gather}
\Lambda'{}_{rs}=-\oint\frac{dz}{2\pi i}\frac{1}{z^{r+1/2}}
\oint\frac{dw}{2\pi i}\frac{1}{w^{s+1/2}}h'(z,w).
\end{gather}

Evaluating the correlation function in the right hand side of
\eqref{def-state-map-h-2-1} one finds
\begin{multline}
\langle \lambda\;\circ\; \gamma'(w)\lambda\;\circ\; \beta'(z) \rangle'
=\langle \left(\frac{\pd\lambda(w)}{\pd w}\right)^{1/2}\gamma'(\lambda(w))
 \left(\frac{\pd\lambda(z)}{\pd z}\right)^{1/2}\beta'(\lambda(z)) \rangle'\\
=\left(\frac{\pd\lambda(w)}{\pd w}\right)^{1/2}
\left(\frac{\pd\lambda(z)}{\pd z}\right)^{1/2}
\langle \eta e^{\phi}(\lambda(w)) e^{-\phi} \pd\xi(\lambda(z)) \rangle'
=\left(\frac{\pd\lambda(w)}{\pd w}\right)^{1/2}
\left(\frac{\pd\lambda(z)}{\pd z}\right)^{1/2}
\frac{1}{\lambda(w)-\lambda(z)}.
\end{multline}
One gets the following answer
\begin{gather}
\Lambda'{}_{rs}=\oint\frac{dz}{2\pi i}\frac{1}{z^{r+1/2}}
\oint\frac{dw}{2\pi i}\frac{1}{w^{s+1/2}}
\left(\frac{\pd\lambda(z)}{\pd z}\right)^{1/2}
\left(\frac{\pd\lambda(w)}{\pd w}\right)^{1/2}
\frac{1}{\lambda(z)-\lambda(w)}.\label{lambda-conf-sliver-1}
\end{gather}

It should be mentioned here that the matrix
\eqref{lambda-conf-sliver-1} coincides with the matrix of the matter
sliver \cite{0112214},\cite{0112231}. This fact was also obtained using
algebraic methods in the previous section.

\subsection{Relationship between star products}
Now we give a relationship between the star-products in the two
theories. We denote these products by $*$ and $*'$ respectively.
\begin{subequations}
\begin{gather}
\langle A|B*C\rangle=\langle f_{1}\circ A(0)f_{2}\circ B(0) f_{3}\circ C(0)\rangle,\\
\langle A|B*'C\rangle=\langle f_{1}\circ A'(0)f_{2}\circ B'(0) f_{3}\circ C'(0)\rangle'.
\end{gather}
\end{subequations}
where $f_{1}(z)=h_{2}^{-1}(h_{3}(z))$, $f_{2}(z)=h_{2}^{-1}(e^{2\pi i/3}h_{3}(z))$,
$f_{3}(z)=h_{2}^{-1}(e^{4\pi i/3}h_{3}(z))$ and
\begin{gather}
h_{n}(z)=\left(\frac{1+iz}{1-iz}\right)^{\frac{2}{n}}
\end{gather}
are the standard
conformal maps used in the definition of the star-product.

The actions of the two theories on a flat world-sheet with
a single defect at the common midpoint of three strings
where we have the deficit of angle of $-\pi$
are related as
\begin{gather}
S'[\phi]=S[\phi]-\frac{1}{2}(\phi(M)+\Bar{\phi}(M)),
\end{gather}
where $M$ denotes the location of the midpoint.
Since the action appears in the path integral through
the combination $e^{-S}$ we have
\begin{gather}
\langle f_{1}\circ A(0)f_{2}\circ B(0) f_{3}\circ C(0)\rangle
=K_{0}\langle f_{1}\circ A'(0)f_{2}\circ B'(0) f_{3}\circ C'(0)\rho^{+}{}'(M)
\rho^{-}{}'(M)\rangle',
\end{gather}
where $K_{0}$ is some normalization constant,
\begin{gather}
\rho^{+}{}'=e^{-\phi/2},\quad \rho^{-}{}'=e^{-\Bar{\phi}/2},
\end{gather}
and $M=f_{1}(i)=f_{2}(i)=f_{3}(i)$. Since in the local coordinate
system the mid-point of the string is at the point $i$, we get
\begin{gather}
f_{1}\circ A'(0)\rho^{+}{}'(M)\rho^{-}{}'(M)=
\lim_{\epsilon\rightarrow 0}|f'_{1}(i+\epsilon)|^{1/4}f_{1}\circ
(A'(0)\rho^{+}{}'(i+\epsilon)\rho^{-}{}'(i+\epsilon)).
\end{gather}
Using the BPZ conjugation $I(z)=-1/z$
\begin{gather}
I\circ(\rho^{+}{}'(i+\epsilon)\rho^{-}{}'(i+\epsilon))\simeq
\rho^{+}{}'(i-\epsilon)\rho^{-}{}'(i-\epsilon)
\end{gather}
we get
\begin{gather}
|B*C\rangle=\lim_{\epsilon\rightarrow 0}|f'_{1}(i+\epsilon)|^{1/4}
\rho^{+}{}'(i-\epsilon)\rho^{-}{}'(i-\epsilon)|B*{}'C\rangle
\propto \rho^{+}{}'(i-\epsilon)\rho^{-}{}'(i-\epsilon)|B*{}'C\rangle.
\end{gather}
Here we omit the possible infinite scale factor since we are analyzing
the solution up to the normalization.

\subsection{Ghost
sliver equation in  twisted CFT }
Here we show that the direct product of the  twisted $(b,c)$ and twisted $(\beta,\gamma)$
slivers
\begin{gather}
|\Xi\rangle=|\Xi'{}_{bc}\rangle\otimes|\Xi'{}_{\beta\gamma}\rangle\label{gh-sliv}
\end{gather}
solves the following equation
\begin{gather}
(c(i)e^{-\phi(i)}-c(-i)e^{-\phi(-i)})|\Xi\rangle+|\Xi*\Xi\rangle=0.
\label{sliv-minus1}
\end{gather}
Using the  $(b,c)$ CFT$'$ \cite{0111129} and $(\beta,\gamma)$ CFT$'$  one rewrites \eqref{sliv-minus1} as
\begin{multline}
(c(i)e^{-\phi(i)}-c(-i)e^{-\phi(-i)})|\Xi'{}_{bc}\rangle\otimes|\Xi'{}_{\beta\gamma}\rangle\\
\propto
-\sigma^{+}{}'(i-\epsilon)\sigma^{-}{}'(i-\epsilon)
\rho^{+}{}'(i-\epsilon)\rho^{-}{}'(i-\epsilon)
|\Xi_{bc}'{}*{}'\Xi_{bc}'{}\rangle\otimes
|\Xi_{\beta\gamma}'{}*{}'\Xi_{\beta\gamma}'{}\rangle\\
\propto
-\sigma^{+}{}'(i-\epsilon)\sigma^{-}{}'(i-\epsilon)
\rho^{+}{}'(i-\epsilon)\rho^{-}{}'(i-\epsilon)
|\Xi_{bc}'\rangle\otimes
|\Xi_{\beta\gamma}'\rangle,
\label{sliv-bcft-eq-minus1}
\end{multline}
where $\sigma^{+}{}'=e^{i\varphi/2}$,  $\sigma^{-}{}'=e^{i\bar{\varphi}/2}$ are conformal operators of
the weight $-1/8$ in the twisted CFT and bosonic ghosts are bosonized as
\begin{gather}
c(z)=e^{i\varphi}(z),\quad b(z)=e^{-i\varphi}(z).
\end{gather}

Let us take the inner product of \eqref{sliv-bcft-eq-minus1} with a Fock space state
$\langle\Phi|$. Using definitions of the sliver
and relation $c(\pm i)=\pm i c{}'(\pm i)$ one gets for the left hand side
\begin{gather}
\langle f\circ \left(\Phi'(0)(c'(i)e^{-\phi(i)}+c'(-i)e^{-\phi(-i)})\right)\rangle'
=\langle f\circ \Phi'(0)\left(c'(i\infty)e^{-\phi(i\infty)}+c'(-i\infty)e^{-\phi(-i\infty)}\right)\rangle'.
\end{gather}
The right hand side of \eqref{sliv-bcft-eq-minus1} is proportional to
\begin{multline}
\langle f\circ \left(\Phi'(0)\sigma^{+}{}'(i+\epsilon)\sigma^{-}{}'(i+\epsilon)
\rho^{+}{}'(i+\epsilon)\rho^{-}{}'(i+\epsilon)\right)\rangle'\\
\propto
\langle f\circ \Phi'(0)\sigma^{+}{}'(i\eta)\sigma^{-}{}'(i\eta)
\rho^{+}{}'(i\eta)\rho^{-}{}'(i\eta)\rangle'\\
\propto
\langle f\circ \Phi'(0)\sigma^{+}{}'(i\eta)\sigma^{+}{}'(-i\eta)
\rho^{+}{}'(i\eta)\rho^{+}{}'(-i\eta)\rangle',
\end{multline}
where $f(i+\epsilon)\simeq \frac{1}{2i}\ln \frac{i\epsilon}{2}\equiv i\eta$
and one has used the Neumann boundary conditions on $\varphi$ and $\phi$ to relate
$\sigma^{-}{}'(i\eta)$ and $\sigma^{+}{}'(-i\eta)$ and
$\rho^{-}{}'(i\eta)$ and $\rho^{+}{}'(-i\eta)$.
Since both correlators are being evaluated on the upper half plane,
the points $\pm i\infty$ correspond to the same points.
The leading terms in the operator product expansion of
$\sigma^{+}{}'(i\eta)$ with $\sigma^{+}{}'(-i\eta)$  and
$\rho^{+}{}'(i\eta)$ with $\rho^{+}{}'(-i\eta)$ are $c(i\infty)$
and $e^{-\phi(i\infty)}$ correspondingly.
So one gets that \eqref{gh-sliv} solves the equation \eqref{sliv-minus1}.

\section{Vacuum Superstring Field Theory Equations.}
\setcounter{equation}{0}

In this section we will describe how the singular VSSFT action
could arise.
A shift of  the SSFT action for both GSO$+$ and GSO$-$
sectors to the tachyon vacuum yields the following action
\begin{multline}
S[\mathcal{A}_{+},\mathcal{A}_{-}]
=\frac{1}{g_{0}^{2}}\left[
\frac{1}{2}\la Y_{-2}|\mathcal{A}_{+},Q_{\textsf{odd}}\mathcal{A}_{+}\ra
+\frac{1}{2}\la Y_{-2}|\mathcal{A}_{-},Q_{\textsf{odd}}\mathcal{A}_{-}\ra
-\la Y_{-2}|\mathcal{A}_{+},Q_{\textsf{even}}\mathcal{A}_{-}\ra\right.\\
\left.+\frac{1}{3}\la Y_{-2}|\mathcal{A}_{+},\mathcal{A}_{+},\mathcal{A}_{+}\ra
-\la Y_{-2}|\mathcal{A}_{+},\mathcal{A}_{-},\mathcal{A}_{-}\ra\right],
\label{action0}
\end{multline}
where $g_{0}$ is an open string coupling constant, $Y_{-2}=Y(i)Y(-i)$, $Y(z)=4c\pd\xi e^{-2\phi}(z)$ is a double step inverse
picture changing operator and $Q_{\textsf{odd}}$
and $Q_{\textsf{even}}$
are
\begin{subequations}
\begin{align}
Q_{\textsf{odd}}Z&=Q_BZ+A_{0,+}* Z-(-1)^{|Z|}Z* A_{0,+},
\\
Q_{\textsf{even}}Z&=A_{0,-}* Z+(-1)^{|Z|}Z* A_{0,-},
\end{align}
\end{subequations}
where $Z$ is a string field in GSO$+$ or GSO$-$ sector, $|Z|$ is a parity of the field $Z$
and $Q_B$ is the BRST charge of superstrings.
$Q_{\textsf{odd}}$ and $Q_{\textsf{even}}$
are expected to be regular since the solution $A_{0,+}$, $A_{0,-}$ describing
the tachyon vacuum is regular. Moreover
$Q_{\textsf{odd}}$ and $Q_{\textsf{even}}$ satisfy the set of axioms \cite{0201197}.
On the other hand it is known that the VSSFT
action is singular. The mechanism describing how this singularity could arise
was proposed by Gaiotto, Rastelli, Sen and Zwiebach \cite{0111129}.
We will describe it below for the case of the action \eqref{action0}.  OSSFT and VSSFT
are related via a singular field redefinition. Let us begin with
$Q_{\textsf{odd}}$ and $Q_{\textsf{even}}$
of the form
\begin{subequations}
\begin{gather}
Q_{\textsf{odd}}=\sum_{r}\int d\sigma\; a_{r}(\sigma)A_{r}(\sigma),\\
Q_{\textsf{even}}=\sum_{r}\int d\sigma\; b_{r}(\sigma)B_{r}(\sigma),
\end{gather}
\end{subequations}
where $a_{r}$ and $b_{r}$ are forms of degree $1-h_{r}$  and $A_r$ and $B_r$
are correspondingly Grassmann odd and even local operators of superghost number one.
We use a double trick, so that $\sigma$ runs from $-\pi$ to $\pi$ and we have only
holomorphic fields. We fix the coordinate system on local patches.
Since $A_{0,+}$ and $A_{0,-}$ are regular we expect $a_{r}(\sigma)$ and $b_{r}(\sigma)$
to be smooth functions of $\sigma$ in local patches \cite{0111129}.

We reparametrize open string coordinate  $\sigma$ to $f(\sigma)$
so that $f(\pi-\sigma)=\pi-f(\sigma)$ for $0\leq\sigma\leq\pi$
and $f(-\sigma-\pi)=-\pi-f(\sigma)$ for $-\pi\leq\sigma\leq0$.
Such reparametrization preserves the star-product, but transforms
$Q_{\textsf{odd}}$ and $Q_{\textsf{even}}$ to
\begin{subequations}
\begin{gather}
Q_{\textsf{odd}}=\sum_{r}\int d\sigma\; a_{r}(\sigma)(f'(\sigma))^{h_r}A'{}_{r}(f(\sigma)),\\
Q_{\textsf{even}}=\sum_{r}\int d\sigma\; b_{r}(\sigma)(f'(\sigma))^{h_r}B'{}_{r}(f(\sigma)).
\end{gather}
\end{subequations}
Consider $f(\sigma)$ squeezed near the midpoint so that
$f'(\pm\pi/2)$ is small and $\int d\sigma\; (f'(\sigma))^{h_r}$, $h_r<0$ gets a large
contribution from a region around the midpoint.
For example one may choose $f'(\sigma)\simeq (\sigma\mp\pi/2)^{2}+\epsilon^{2}$
for $\sigma\simeq\pm\pi/2$. One gets that
$Q_{\textsf{odd}}$ and $Q_{\textsf{even}}$
get a dominant contribution from the lowest dimension operators $c$ and $\gamma$
\begin{subequations}
\begin{gather}
Q_{\textsf{odd}}=\epsilon^{-1}(c(i)-c(-i)),\\
Q_{\textsf{even}}=\epsilon^{-1/2}(\gamma(i)-\gamma(-i)),
\end{gather}
\end{subequations}
Now we could make a singular field redefinition
$\mathcal{A}_{+}=\epsilon^{-1}\mathrm{A}_{+}$,
$Q_{\textsf{odd}}=\epsilon^{-1}\mathcal{Q}_{\textsf{odd}}$,
$\mathcal{A}_{-}=\epsilon^{-1}\mathrm{A}_{-}$ and
$Q_{\textsf{even}}=\epsilon^{-1}\mathcal{Q}_{\textsf{even}}$.
Since $\mathcal{Q}_{\textsf{odd}}=c(i)-c(-i)$ and
$\mathcal{Q}_{\textsf{even}}=\epsilon^{1/2}(\gamma(i)-\gamma(-i))$ this choice
of kinetic operators satisfies axioms \cite{0201197}.
Actually we get that $\mathcal{Q}_{\textsf{even}}\rightarrow 0$ as $\epsilon\rightarrow 0$.
The action \eqref{action0} takes the form
\begin{multline}
S[\mathrm{A}_{+},\mathrm{A}_{-}]
=\frac{\kappa_{0}}{g_{0}^{2}}\left[
\frac{1}{2}\la Y_{-2}|\mathrm{A}_{+},\mathcal{Q}_{\textsf{odd}}\mathrm{A}_{+}\ra
+\frac{1}{2}\la Y_{-2}|\mathrm{A}_{-},\mathcal{Q}_{\textsf{odd}}\mathrm{A}_{-}\ra\right.\\
\left.+\frac{1}{3}\la Y_{-2}|\mathrm{A}_{+},\mathrm{A}_{+},\mathrm{A}_{+}\ra
-\la Y_{-2}|\mathcal{A}_{+},\mathrm{A}_{-},\mathrm{A}_{-}\ra\right],
\label{action1}
\end{multline}
where $\kappa_{0}=\epsilon^{-3}$. The equations of motion for this action are
\begin{subequations}
\begin{gather}
\mathcal{Q}_{\textsf{odd}}\mathrm{A}_{+}+\mathrm{A}_{+}*\mathrm{A}_{+}
-\mathrm{A}_{-}*\mathrm{A}_{-}=0,\\
\mathcal{Q}_{\textsf{odd}}\mathrm{A}_{-}+\mathrm{A}_{+}*\mathrm{A}_{-}
-\mathrm{A}_{-}*\mathrm{A}_{+}=0.
\end{gather}
\end{subequations}

However to get non-trivial solution for $\mathrm{A}_{-}$
it is convenient to make another
field redefinition: $\mathcal{A}_{+}=\epsilon^{-1}\mathrm{A}_{+}$,
$Q_{\textsf{odd}}=\epsilon^{-1}\mathcal{Q}_{\textsf{odd}}$,
$\mathcal{A}\underline{}_{-}=\epsilon^{-1/2}\mathrm{A}'{}_{-}$ and
$Q_{\textsf{even}}=\epsilon^{-1/2}\mathcal{Q}'{}_{\textsf{even}}$.
One gets the following singular action
\begin{multline}
S[\mathrm{A}_{+},\mathrm{A}'{}_{-}]
=\frac{\kappa_{0}}{g_{0}^{2}}\left[
\frac{1}{2}\la Y_{-2}|\mathrm{A}_{+},\mathcal{Q}_{\textsf{odd}}\mathrm{A}_{+}\ra
+\frac{1}{3}\la Y_{-2}|\mathrm{A}_{+},\mathrm{A}_{+},\mathrm{A}_{+}\ra\right]\\
+\frac{\kappa'{}_{0}}{g_{0}^{2}}\left[
\frac{1}{2}\la Y_{-2}|\mathrm{A}'{}_{-},\mathcal{Q}_{\textsf{odd}}\mathrm{A}'{}_{-}\ra
-\la Y_{-2}|\mathrm{A}_{+},\mathcal{Q}'{}_{\textsf{even}}\mathrm{A}'{}_{-}\ra
-\la Y_{-2}|\mathrm{A}_{+},\mathrm{A}'{}_{-},\mathrm{A}'{}_{-}\ra\right],
\label{action-kappa}
\end{multline}
where $\kappa'{}_{0}=\epsilon^{-2}$.

 The gauge invariance in terms of redefined fields and charges is
\begin{subequations}
\begin{gather}
\delta \mathrm{A}_{+}=\mathcal{Q}_{\textsf{odd}}\Lambda_{+}+[\mathrm{A}_{+},\Lambda_{+}]
+\epsilon^{1/2}(\mathcal{Q}'{}_{\textsf{even}}\Lambda_{-}+[\mathrm{A}'{}_{-},\Lambda_{-}]),\\
\delta \mathrm{A}'{}_{-}=\mathcal{Q}'{}_{\textsf{even}}\Lambda_{+}+[\mathrm{A}'{}_{-},\Lambda_{+}]
+\epsilon^{-1/2}(\mathcal{Q}_{\textsf{odd}}\Lambda_{-}+[\mathrm{A}_{+},\Lambda_{-}]).
\end{gather}
\end{subequations}
This action gives the following equations of motion
\begin{subequations}
\begin{gather}
\mathcal{Q}_{odd}\mathrm{A}_{+}+\mathrm{A}_{+}*\mathrm{A}_{+}-
\epsilon^{-1}(\mathcal{Q}'{}_{even}\mathrm{A}'{}_{-}+\mathrm{A}'{}_{-}*\mathrm{A}'{}_{-})=0,\\
\mathcal{Q}_{odd}\mathrm{A}'{}_{-}+\mathrm{A}_{+}*\mathrm{A}'{}_{-}-
\mathcal{Q}'{}_{even}\mathrm{A}_{+}-\mathrm{A}'{}_{-}*\mathrm{A}_{+}=0.
\end{gather}
\end{subequations}
Here we omit $Y_{-2}$. We will solve them in the factorized form
\begin{subequations}
\begin{gather}
\mathcal{Q}_{odd}\mathrm{A}_{+}+\mathrm{A}_{+}*\mathrm{A}_{+}=0,\label{eq1}\\
\mathcal{Q}'{}_{even}\mathrm{A}'{}_{-}+\mathrm{A}'{}_{-}*\mathrm{A}'{}_{-}=0,\label{eq2}\\
\mathcal{Q}_{odd}\mathrm{A}'{}_{-}+\mathrm{A}_{+}*\mathrm{A}'{}_{-}=0,\label{eq3}\\
\mathcal{Q}'{}_{even}\mathrm{A}_{+}+\mathrm{A}'{}_{-}*\mathrm{A}_{+}=0.\label{eq4}
\end{gather}
\end{subequations}
Equation \eqref{eq1} is the equation for the twisted $(b,c)$ sliver.
The fermionic ghost part of $\mathrm{A}_{+}$ should be chosen as a conformal
sliver, i.e. the
sliver in the zero picture, that satisfies the projector equation.

\subsection{Ghost sliver equations}
The surface state corresponding to the conformal
map $f(\xi)$ with insertion $\phi_{d}(z)$ \cite{RZ},\cite{0201095} is defined as
\begin{gather}
U_{f}^{\dag}\phi_{d}(z)|0\rangle,
\end{gather}
where $f\circ\phi_{d}(z)\equiv [f'(z)]^{d}\phi'{}_{d}(f(z))=U_{f}\phi_{d}(z)U^{-1}_{f}$ for the field $\phi_{d}(z)$ of the
weight $d$ and $U_{f}^{\dag}=U^{-1}_{I\circ f\circ I}$.

One obtains
\begin{gather}
U_{f}^{\dag}\,I\circ f\circ I\circ\phi_{d}(z)|0\rangle=
\phi_{d}(z)U_{f}^{\dag}|0\rangle.
\end{gather}

In particular for the map $f(\xi)=\arctan\xi$ corresponding to the sliver and the field $\phi_{d}(z)$
of the weight $d=0$
one gets for $z=\pm i\pm\epsilon$, $\epsilon\rightarrow 0$
\begin{gather}
I\circ f\circ I\circ\phi_{0}(\pm i\mp\epsilon)=\phi_{0}(\pm i\eta^{-1}),
\end{gather}
since
\begin{gather}
f(\pm i\pm\epsilon)=\arctan(\pm i\pm\epsilon)=\frac{\pm1}{2i}\ln\frac{i\epsilon}{2}=\pm i\eta,\quad \eta\rightarrow\infty.
\end{gather}

Let us consider equation \eqref{eq2}
\begin{gather}
Y(i)Y(-i)[(\gamma(i)-\gamma(-i))\mathrm{A}'{}_{-}+\mathrm{A}'{}_{-}*\mathrm{A}'{}_{-}]=0.
\end{gather}
Let us choose $\mathrm{A}'{}_{-}$ as a sliver with insertion $\gamma(0)$
\begin{gather}
\mathrm{A}'{}_{-}=U_{f}^{\dag}\,\gamma(0)|0\rangle.
\end{gather}
Changing the picture one finds
\begin{gather}
Y(\pm i)\mathrm{A}'{}_{-}=U_{f}^{\dag}\,Y(0)\gamma(0)|0\rangle
=-4\,U_{f}^{\dag}\,c(0)e^{-\phi(0)}|0\rangle\equiv\mathrm{A}_{-}^{-}.
\end{gather}
Ghost part of $\mathrm{A}_{-}^{-}$ is direct product of
the sliver in the minus one picture for the fermionic ghosts
and the twisted sliver for the bosonic ghosts.
So we get the following equation
\begin{gather}
(c(i)e^{-\phi(i)}-c(-i)e^{-\phi(-i)})\mathrm{A}_{-}^{-}+\mathrm{A}_{-}^{-}*\mathrm{A}_{-}^{-}=0
\end{gather}
This is indeed the equation for the ghost sliver that we have solved in Section 3.
Equations \eqref{eq3} and \eqref{eq4} are also satisfied for such $\mathrm{A}_{+}$
and $\mathrm{A}'{}_{-}$.

\section{Conclusion and Discussions.}

In this paper Neveu-Schwarz ghost slivers in  pictures zero and minus one are constructed.
It is shown that the sliver in a twisted CFT is in fact a projector
with respect to the twisted star product and
that this sliver multiplied by $(b,c)$ twisted sliver and other slivers
solve super VSFT equations.

It is worth to note that
our results can be used in the following directions.
It is interesting to consider excitations on a solution
of \eqref{int-sliv-minus1} to check if one can reproduce the
standard perturbative spectrum of NS string.
In spite of the fact that we consider only cubic superstring field
theory, our results
can be applied to the Berkovits non-polynomial superstring field theory.

Let us note that besides operator and CFT formalism
string star algebra admits a nice half-string description
\cite{0105058,GT01,GT02,0106157,0202030,0202087}
(see \cite{0111208} for a review and refs. therein) and it would be
interesting to generalize
this formalizm to superstrings.

One of the unsolved interesting problems
is the construction of the analog of the continuous Moyal product
for the fermionic star algebra in the framework of \cite{0202087}.


\section*{Acknowledgments}

We would like to thank Dmitri Belov for many useful discussions.
This work was supported in part
by RFBR grant  02-01-00695  and RFBR grant for leading scientific
schools and  by INTAS grant
99-0590.

\appendix
\section*{Appendix}
\addcontentsline{toc}{section}{Appendix}
\renewcommand {\theequation}{\thesection.\arabic{equation}}

\section{$\beta, \gamma$ vertices.}
\setcounter{equation}{0}

In this section we give an algebraic construction
of the fermionic ghost vertices: identity, reflector and
three string vertex \cite{GrJe3} over the vacuum in the $-1$ picture.
Let us first solve the overlap equations
for the identity state
\begin{subequations}
\begin{gather}
[\beta_{\pm}(\sigma)\pm i\beta_{\pm}(\pi-\sigma)]|I^{\beta\gamma}\rangle=0,\quad
 \sigma\in(0,\pi/2),\\
[\gamma_{\pm}(\sigma)\pm i\gamma_{\pm}(\pi-\sigma)]|I^{\beta\gamma}\rangle=0,\quad
 \sigma\in(0,\pi/2).
\end{gather}
\end{subequations}
Using the identity one can rewrite
these overlaps in terms of $\beta_{+}(\sigma)$ and
$\gamma_{+}(\sigma)$, which are defined on the whole interval $(-\pi,\pi)$
\begin{gather}
\beta_{+}(\sigma)=\beta_{-}(-\sigma),\quad
\gamma_{+}(\sigma)=\gamma_{-}(-\sigma),
\quad \sigma\in[-\pi,0],
\end{gather}
in the following form
\begin{subequations}
\begin{gather}
\beta_{+}(\sigma)=
  \begin{cases}
    -i\beta_{+}(\pi-\sigma),\quad (0,\pi/2), \\
    i\beta_{+}(-\pi-\sigma),\quad (-\pi/2,0), \\
    i\beta_{+}(\pi-\sigma),\quad (\pi/2,\pi), \\
    -i\beta_{+}(-\pi-\sigma),\quad (-\pi,-\pi/2);
  \end{cases}\\
  \gamma_{+}(\sigma)=
  \begin{cases}
    -i\gamma_{+}(\pi-\sigma),\quad (0,\pi/2), \\
    i\gamma_{+}(-\pi-\sigma),\quad (-\pi/2,0), \\
    i\gamma_{+}(\pi-\sigma),\quad (\pi/2,\pi), \\
    -i\gamma_{+}(-\pi-\sigma),\quad (-\pi,-\pi/2).
  \end{cases}
\end{gather}
\end{subequations}
 Integration of these equations via the relations
$\frac{1}{2\pi}\int_{-\pi}^{\pi}d\sigma
e^{-ir\sigma}\beta_{+}(\sigma)=\beta_{r}$
and $\frac{1}{2\pi}\int_{-\pi}^{\pi}d\sigma
e^{-ir\sigma}\gamma_{+}(\sigma)=\gamma_{r}$ yields the following
overlap equations for the identity state
\begin{subequations}
\begin{gather}
\beta_{r}=-F_{rs}\beta_{s}-\tilde{F}_{rs}\beta_{-s},\\
\beta_{-r}=\tilde{F}_{rs}\beta_{s}+F_{rs}\beta_{-s},\\
\gamma_{r}=-F_{rs}\gamma_{s}-\tilde{F}_{rs}\gamma_{-s},\\
\gamma_{-r}=\tilde{F}_{rs}\gamma_{s}+F_{rs}\gamma_{-s}.
\end{gather}
\end{subequations}
Here $r,s\geq\frac{1}{2}$ and the hermitian matrices $F_{rs}$ and
$\tilde{F}_{rs}$ are the same as for the matter fermionic case \cite{GrJe3}
\begin{subequations}
\begin{gather}
F_{rs}=-\frac{2}{\pi}\frac{\imath^{r-s}}{r+s},\quad
r=s\text{ mod}(2),\\
\tilde{F}_{rs}=\frac{2}{\pi}\frac{\imath^{r+s}}{s-r},\quad
r=s+1\text{ mod}(2).
\end{gather}
\end{subequations}
They have  the following properties
\begin{gather}
F^{2}-\tilde{F}^{2}=1,\quad [F,\tilde{F}]=0,\\
CFC=-F,\quad F^{T}=F,\quad C\tilde{F}C=\tilde{F},\quad
\tilde{F}^{T}=-\tilde{F},
\end{gather}
where $C_{rs}=(-1)^{r}\delta_{rs}$.
It is reasonable to search for  the identity state in the following form
\begin{gather}
|I^{\beta\gamma}\rangle=\exp(\beta_{-r}I^{\beta\gamma}_{rs}\gamma_{-s})|-1\rangle,
\end{gather}
where $|-1\rangle$ is the vacuum in the minus one picture, which is
annihilated by $\beta_{r}$, $\gamma_{r}$ for $r\geq1/2$.
One gets the following expressions for the matrix $I^{\beta\gamma}_{rs}$
\begin{gather}
I^{\beta\gamma}=-\frac{\tilde{F}}{1+F}=\frac{1-F}{\tilde{F}}=I^{-1},
\end{gather}
where $I$ is the identity of the matter fermions.
Expression for reflector is found to be
\begin{equation}
|R^{\beta\gamma}\rangle=\exp(i\beta^{1}_{-r} (-1)^{r}\gamma^{2}_{-r}
-i\beta^{2}_{-r}(-1)^{r}\gamma^{1}_{-r})|-1\rangle_{12}.\label{refl}
\end{equation}
For the three string fermionic ghost vertex one gets the
following overlap equations
\begin{subequations}
\begin{gather}
[\beta^{a}_{\pm}(\sigma)\pm i\beta^{a-1}_{\pm}(\pi-\sigma)]|V^{\beta\gamma}\rangle=0,\quad (0,\pi/2),\\
[\gamma^{a}_{\pm}(\sigma)\pm i\gamma^{a-1}_{\pm}(\pi-\sigma)]|V^{\beta\gamma}\rangle=0,\quad (0,\pi/2).
\end{gather}
\end{subequations}
These overlap equations can be explicitly solved in terms of the new
$\Zh_{3}$ Fourier variables
\begin{subequations}
\begin{gather}
\mathcal{B}^{1}=\frac{1}{\sqrt{3}}(\beta^{1}_{+}+\beta^{2}_{+}+\beta^{3}_{+}),\\
\mathcal{B}^{2}=\frac{1}{\sqrt{3}}(\beta^{1}_{+}+\alpha\beta^{2}_{+}+\alpha^{*}\beta^{3}_{+})\equiv \mathcal{B},\\
\mathcal{B}^{3}=\frac{1}{\sqrt{3}}(\beta^{1}_{+}+\alpha^{*}\beta^{2}_{+}+\alpha\beta^{3}_{+})\equiv\Bar{\mathcal{B}};\\
\mathcal{G}^{1}=\frac{1}{\sqrt{3}}(\gamma^{1}_{+}+\gamma^{2}_{+}+\gamma^{3}_{+}),\\
\mathcal{G}^{2}=\frac{1}{\sqrt{3}}(\gamma^{1}_{+}+\alpha\gamma^{2}_{+}+\alpha^{*}\gamma^{3}_{+})\equiv \mathcal{G},\\
\mathcal{G}^{3}=\frac{1}{\sqrt{3}}(\gamma^{1}_{+}+\alpha^{*}\gamma^{2}_{+}+\alpha\gamma^{3}_{+})\equiv\Bar{\mathcal{G}}.
\end{gather}
\end{subequations}
In terms of these variables one gets the
following overlaps
\begin{subequations}
\begin{gather}
\mathcal{B}^{1}(\sigma)=
\begin{cases}
-\imath \mathcal{B}^{1}(\pi-\sigma),\quad |\sigma|\leq\frac{\pi}{2},\\
\imath \mathcal{B}^{1}(\pi-\sigma),\quad\frac{\pi}{2}\leq|\sigma|\leq\pi,
\end{cases}\\
\mathcal{B}^{2}(\sigma)=
\begin{cases}
-\imath\alpha \mathcal{B}^{2}(\pi-\sigma),\quad |\sigma|\leq\frac{\pi}{2},\\
\imath\alpha^{*}\mathcal{B}^{2}(\pi-\sigma),\quad\frac{\pi}{2}\leq|\sigma|\leq\pi,
\end{cases}\\
\mathcal{B}^{3}(\sigma)=
\begin{cases}
-\imath\alpha^{*}\mathcal{B}^{3}(\pi-\sigma),\quad |\sigma|\leq\frac{\pi}{2},\\
\imath\alpha
\mathcal{B}^{3}(\pi-\sigma),\quad\frac{\pi}{2}\leq|\sigma|\leq\pi,
\end{cases}
\end{gather}
\end{subequations}
and analogous  ones  for  the  variables $\mathcal{G}(\sigma)$.
All these overlaps differ from the overlaps for the matter fermions
only by the sign in the right hand side. The overlap equations for
$\mathcal{B}(\sigma)$ can be rewritten  in components as
\begin{subequations}
\begin{gather}
\mathcal{B}_{r}=\frac{1}{2}F_{rs}\mathcal{B}_{s}+\frac{1}{2}(\tilde{F}_{rs}-\sqrt{3}C)\mathcal{B}_{-s},\\
\mathcal{B}_{-r}=-\frac{1}{2}(\tilde{F}_{rs}-\sqrt{3}C)\mathcal{B}_{s}-\frac{1}{2}F_{rs}\mathcal{B}_{-s}.
\end{gather}
\end{subequations}
and analogous for variables $\mathcal{G}$.
We search for the three string vertex of the
following form in terms of the new variables
\begin{gather}
|V^{\beta\gamma}\rangle=
\exp(\mathcal{B}^{1}_{-r}I^{\beta\gamma}_{rs}\mathcal{G}^{1}_{-s}+
\Bar{\mathcal{B}}_{-r}U^{\beta\gamma}_{rs}\mathcal{G}_{-s}
+\mathcal{B}_{-r}\tilde{U}^{\beta\gamma}_{rs}\Bar{\mathcal{G}}_{-s})|-1\rangle_{123}.
\end{gather}
By solving the overlaps for $\mathcal{B}$ and $\mathcal{G}$ one gets
the following expressions for $U^{\beta\gamma}$ and
$\tilde{U}^{\beta\gamma\,T}$
\begin{subequations}
\begin{gather}
U^{\beta\gamma}=\tilde{U}^{\beta\gamma\,T}=\frac{\tilde{F}-\sqrt{3}C}{2-F}=U^{-1},\\
U^{\beta\gamma}=\tilde{U}^{\beta\gamma\,T}=-\frac{2+F}{\tilde{F}-\sqrt{3}C}=U^{-1},
\end{gather}
\end{subequations}
where $U$ is the corresponding matrix  for the matter fermions.
Solving the overlaps for $\Bar{\mathcal{B}}$ and $\Bar{\mathcal{G}}$ one gets the following
\begin{subequations}
\begin{gather}
U^{\beta\gamma\,T}=\tilde{U}^{\beta\gamma}=\frac{\tilde{F}+\sqrt{3}C}{2-F}=CU^{\beta\gamma}C,\\
U^{\beta\gamma\,T}=\tilde{U}^{\beta\gamma}=-\frac{2+F}{\tilde{F}+\sqrt{3}C}=CU^{\beta\gamma}C.
\end{gather}
\end{subequations}
Rewriting the three string fermionic ghost vertex in terms of the old variables
one gets
\begin{gather}
|V^{\beta\gamma}\rangle=
\exp(\beta^{a}_{-r}V^{\beta\gamma,\,ab}_{rs}\gamma^{b}_{-s})|-1\rangle_{123},
\end{gather}
where
\begin{subequations}
\begin{gather}
V^{\beta\gamma,\,a\,a}=\frac{1}{3}(I^{\beta\gamma}+U^{\beta\gamma}+CU^{\beta\gamma}C),\\
V^{\beta\gamma,\,a\,a+1}=\frac{1}{3}(I^{\beta\gamma}+\alpha U^{\beta\gamma}+\alpha^{*} CU^{\beta\gamma}C),\\
V^{\beta\gamma,\,a\,a-1}=\frac{1}{3}(I^{\beta\gamma}+\alpha^{*}
U^{\beta\gamma}+\alpha CU^{\beta\gamma}C).
\end{gather}
\end{subequations}
More explicitly one  obtains  the following expressions
\begin{subequations}
\label{zero-vert}
\begin{gather}
V^{\beta\gamma,\,11}=\frac{F\tilde{F}}{(1+F)(2-F)},\\
V^{\beta\gamma,\,12}=\frac{-\tilde{F}-i C(1+F)}{(1+F)(2-F)},\\
V^{\beta\gamma,\,21}=\frac{-\tilde{F}+i C(1+F)}{(1+F)(2-F)}.
\end{gather}
\end{subequations}
We should mention here that the fermionic ghost vertices differs from
that of the fermionic matter only by the signs of the matrices $F$, $\tilde{F}$, $C$.
Changing the signs of these matrices one gets the matter fermionic vertices.
Using the following notation
\begin{equation}
M^{\beta\gamma}_{ab}=CV^{\beta\gamma,\,ab},
\end{equation}
one gets the following properties for $M^{\beta\gamma}_{ab}$
\begin{subequations}
\label{matrix-gh-prop}
\begin{gather}
M^{\beta\gamma}_{12}+M^{\beta\gamma}_{21}+M^{\beta\gamma}_{11}=CI^{\beta\gamma},\\
[M^{\beta\gamma}_{ab},M^{\beta\gamma}_{cd}]=0,\quad \forall \;\; a, b, c, d,\\
M^{\beta\gamma}_{12}M^{\beta\gamma}_{21}=M^{\beta\gamma\,2}_{11}-CI^{\beta\gamma\,-1}M^{\beta\gamma}_{11}.
\end{gather}
\end{subequations}

\section{Equivalence of algebraic and conformal solutions.}
\setcounter{equation}{0}

Using the results on the spectroscopy of the
NS star algebra \cite{0112231} we generalize the technique of Okuyama \cite{0201015},\cite{0201149}
to the NS case and prove the equivalence
of the algebraic and conformal definitions of the NS matter sliver and NS ghost sliver in the
minus one picture.

Matrices $F$ and $C\tilde{F}$ have the spectrum
\begin{gather}
\phi(\kappa)=-\frac{1}{\cosh(\frac{\pi\kappa}{2})},\quad
\tilde{\phi}(\kappa)=i\tanh(\frac{\pi\kappa}{2})
\end{gather}
correspondingly. The corresponding generation function of eigenvectors is given by
\begin{gather}
f_{w^{(\kappa)}}(z)=w^{(\kappa)}_{\frac{1}{2}}\frac{z}{\sqrt{1+z^2}}\exp(-\kappa\arctan(z)).
\label{eigfun}
\end{gather}
The spectrum of the NS matter sliver is given by
\begin{gather}
T(\kappa)=
  \begin{cases}
    i\frac{\kappa}{|\kappa|}e^{-\pi|\kappa|/2} & \kappa\neq0, \\
    0 & \kappa=0.
  \end{cases}
\end{gather}
Let us denote as $|z\rangle$ and $|\kappa\rangle$ the infinite dimensional vectors
$|z\rangle=(z,z^{2},z^{3},\dots)^{T}$ and $|\kappa\rangle=(w^{(\kappa)}_{1},w^{(\kappa)}_{2},w^{(\kappa)}_{3},\dots)^{T}$.
The generating function \eqref{eigfun} of eigenvectors in these notations is given by
\begin{gather}
f_{w^{(\kappa)}}(z)=w^{(\kappa)}_{\frac{1}{2}}\frac{z}{\sqrt{1+z^{2}}}\exp(-\kappa\tan^{-1}z)
=\sum_{n=1}^{\infty}w_{n}^{(\kappa)}z^{n}\equiv\langle z|w^{(\kappa)}\rangle.
\end{gather}
The inner product of two vectors $|w\rangle$ and $|w'\rangle$ is defined as
\begin{gather}
\langle w|w'\rangle\equiv\sum_{n=1}^{\infty}w_{n}w'{}_{n}=\int_{-\pi}^{\pi}\frac{d\theta}{2\pi}
\langle w|e^{i\theta}\rangle\langle e^{-i\theta}|w'\rangle.
\end{gather}
Let us compute the inner product of two eigenvectors
\begin{multline}
\langle \kappa|p\rangle=
\int_{-\pi/2}^{3\pi/2}\frac{d\theta}{2\pi}f^{*}_{w^{(\kappa)}}(e^{i\theta})
f_{w{}^{(p)}}(e^{i\theta})=
\int_{-\pi/2}^{3\pi/2}\frac{d\theta}{2\pi}
\langle \kappa|e^{i\theta}\rangle\langle e^{-i\theta}|p\rangle\\
=\int_{-\pi/2}^{3\pi/2}\frac{d\theta}{2\pi}\frac{e^{i\theta}}{\sqrt{1+e^{2i\theta}}}\frac{e^{-i\theta}}{\sqrt{1+e^{-2i\theta}}}
e^{-\kappa \tan^{-1}e^{i\theta}}e^{-p \tan^{-1}e^{-i\theta}}.
\end{multline}
We specify the branch of the function
\begin{gather}
\tan^{-1}z=\frac{1}{2i}\log\frac{1+iz}{1-iz},\quad \tan^{-1}0=0.
\end{gather}
We change variables
\begin{subequations}
\begin{gather}
\tan^{-1}e^{i\theta}=\frac{\pi}{4}+ix,\quad
\tan^{-1}e^{-i\theta}=\frac{\pi}{4}-ix,\quad
\tan\frac{\theta}{2}=\tanh x,\quad \left[-\frac{\pi}{2}\leq\theta\leq\frac{\pi}{2}\right]\\
\tan^{-1}e^{i\theta}=-\frac{\pi}{4}-ix,\quad
\tan^{-1}e^{-i\theta}=-\frac{\pi}{4}+ix,\quad
-\cot\frac{\theta}{2}=\tanh x.\quad \left[\frac{\pi}{2}\leq\theta\leq\frac{3\pi}{2}\right]
\end{gather}
\end{subequations}
One finds
\begin{multline}
\langle \kappa|p\rangle=\int_{-\infty}^{\infty}\frac{dx}{2\pi}
e^{-\pi\kappa/4-i\kappa x}e^{-\pi p/4+ipx}
+\int_{-\infty}^{\infty}\frac{dx}{2\pi}
e^{\pi\kappa/4+i\kappa x}e^{\pi p/4-ipx}\\
=e^{-\pi\kappa/2}\delta(\kappa-p)+e^{\pi\kappa/2}\delta(\kappa-p)
=2\cosh\left(\frac{\pi\kappa}{2}\right)\delta(\kappa-p)
\equiv \mathcal{N}(\kappa)\delta(\kappa-p).
\end{multline}
Introducing the normalized eigenvectors $|\hat{\kappa}\rangle=\mathcal{N}(\kappa)^{-1/2}|\kappa\rangle$
we get the completeness relation
\begin{gather}
1=\int_{-\infty}^{\infty}d\kappa\; |\hat{\kappa}\rangle\langle\hat{\kappa}|=
\int_{-\infty}^{\infty}d\kappa\;\mathcal{N}(\kappa)\;|\kappa\rangle\langle\kappa|.
\end{gather}
It is a good check to consider the following relation
\begin{multline}
\langle z|w\rangle=\sum_{n=1}^{\infty}z^{n}w^{n}=\frac{zw}{1-zw}
=\int_{-\infty}^{\infty}d\kappa \langle z|\hat{\kappa}\rangle\langle\hat{\kappa}|w\rangle=\\
=\frac{z}{\sqrt{1+z^{2}}}\frac{w}{\sqrt{1+w^{2}}}
\int_{-\infty}^{\infty}d\kappa \frac{1}{2\cosh(\frac{\pi\kappa}{2})}e^{-\kappa \tan^{-1}z}
e^{-\kappa \tan^{-1}w}\\
=\frac{z}{\sqrt{1+z^{2}}}\frac{w}{\sqrt{1+w^{2}}}
\int_{0}^{\infty}d\kappa \frac{1}{\cosh(\frac{\pi\kappa}{2})}\cosh(-\kappa (\tan^{-1}z+\tan^{-1}w))\\
=\frac{z}{\sqrt{1+z^{2}}}\frac{w}{\sqrt{1+w^{2}}}
\frac{\sqrt{1+z^2}\sqrt{1+w^2}}{1-zw}=\frac{zw}{1-zw}.
\end{multline}
Here we have used
\begin{subequations}
\begin{gather}
\int_{0}^{\infty}dx\;\frac{\sinh(ax)}{\sinh(bx)}=\frac{\pi}{2b}\tan\frac{\pi a}{2b},\quad (|\Re a|<|\Re b|),\\
\int_{0}^{\infty}dx\;\frac{\cosh(ax)}{\cosh(bx)}=\frac{\pi}{2b}\frac{1}{\cos\frac{\pi a}{2b}},\quad (|\Re a|<|\Re b|).
\end{gather}
\end{subequations}

Further we show the equivalence of the algebraic and conformal sliver.
The generating function of the matter NS sliver $S_{rs}$ is given by
\begin{multline}
h^{\mu\nu}(z,w)\equiv\langle\Xi|\psi^{\mu}(w)\psi^{\nu}(z)|0\rangle
=\langle 0|\exp(-\frac{1}{2}\psi_{r}S_{rs}\psi_{s})\psi^{\mu}(w)\psi^{\nu}(z)|0\rangle\\
=-\frac{1}{2}w^{r-1/2}z^{s-1/2}S_{rs}\eta^{\mu\nu}-\frac{1}{2}\eta^{\mu\nu}\frac{1}{w-z}.
\end{multline}
On the other hand the conformal definition gives
\begin{gather}
h^{\mu\nu}(z,w)\equiv\langle f\circ\psi^{\mu}(w)f\circ\psi^{\nu}(z)\rangle=\left(\frac{\pd f(w)}{\pd w}\right)^{1/2}\left(\frac{\pd f(z)}{\pd z}\right)^{1/2}
\left(-\frac{1}{2}\right)\frac{\eta^{\mu\nu}}{f(w)-f(z)}.
\end{gather}
Substituting the conformal map $f(z)=\tan^{-1}z$ for the sliver
we get the following equation that should hold
\begin{gather}
\langle z|S|w\rangle=\frac{zw}{z-w}+\frac{z}{\sqrt{1+z^{2}}}\frac{w}{\sqrt{1+w^{2}}}\frac{1}{\tan^{-1}w-\tan^{-1}z}.
\end{gather}
Indeed
\begin{multline}
\langle -z|S|w\rangle=\langle z|iCS|w\rangle=\frac{z}{\sqrt{1+z^{2}}}
\frac{w}{\sqrt{1+w^{2}}}\int_{0}^{\infty}d\kappa\;e^{-\frac{\pi\kappa}{2}}
\frac{\sinh(\kappa(\tan^{-1}z+\tan^{-1}w))}{\cosh(\frac{\pi\kappa}{2})}\\
=\frac{z}{\sqrt{1+z^{2}}}\frac{w}{\sqrt{1+w^{2}}}
\left[\frac{1}{\sin(\tan^{-1}z+\tan^{-1}w)}-\frac{1}{\tan^{-1}z+\tan^{-1}w}\right]\\
=\frac{zw}{z+w}-\frac{z}{\sqrt{1+z^{2}}}\frac{w}{\sqrt{1+w^{2}}}\frac{1}{\tan^{-1}z+\tan^{-1}w}.
\end{multline}
Here we have used
\begin{gather}
\int_{0}^{\infty}dx\frac{\sinh(ax)}{e^{bx}+1}=\frac{\pi}{2b}\frac{1}{\sin(\frac{a\pi}{b})}-\frac{1}{2a},\quad p>a,\; p>0.
\end{gather}
The formulae for the fermionic ghosts in the minus one picture are straightforward.
The generating function of the fermionic ghost sliver $S^{\beta\gamma}$ in the minus one picture is given by
\begin{multline}
h^{\mu\nu}(z,w)\equiv\langle\Xi|\gamma(w)\beta(z)|-1\rangle
=\langle -1|\exp(\gamma_{r}S^{\beta\gamma}_{rs}\beta_{s})\gamma(w)\beta(z)|-1\rangle\\
=w^{r-1/2}z^{s-1/2}S^{\beta\gamma}_{rs}+\frac{1}{w-z}.
\end{multline}
On the other hand the conformal definition gives
\begin{gather}
h^{\mu\nu}(z,w)\equiv\langle f\circ\gamma(w)f\circ\beta(z)\rangle=\left(\frac{\pd f(w)}{\pd w}\right)^{1/2}\left(\frac{\pd f(z)}{\pd z}\right)^{1/2}
\frac{1}{f(w)-f(z)}.
\end{gather}
Substituting the conformal map $f(z)=\tan^{-1}z$ for the sliver
we get the following equation that should hold
\begin{gather}
\langle z|S^{\beta\gamma}|w\rangle=\frac{zw}{z-w}+\frac{z}{\sqrt{1+z^{2}}}\frac{w}{\sqrt{1+w^{2}}}\frac{1}{\tan^{-1}w-\tan^{-1}z}.
\end{gather}
Indeed
\begin{multline}
\langle -z|S^{\beta\gamma}|w\rangle=\langle z|iCS^{\beta\gamma}|w\rangle=\frac{z}{\sqrt{1+z^{2}}}
\frac{w}{\sqrt{1+w^{2}}}\int_{0}^{\infty}d\kappa\;e^{-\frac{\pi\kappa}{2}}
\frac{\sinh(\kappa(\tan^{-1}z+\tan^{-1}w))}{\cosh(\frac{\pi\kappa}{2})}\\
=\frac{z}{\sqrt{1+z^{2}}}\frac{w}{\sqrt{1+w^{2}}}
\left[\frac{1}{\sin(\tan^{-1}z+\tan^{-1}w)}-\frac{1}{\tan^{-1}z+\tan^{-1}w}\right]\\
=\frac{zw}{z+w}-\frac{z}{\sqrt{1+z^{2}}}\frac{w}{\sqrt{1+w^{2}}}\frac{1}{\tan^{-1}z+\tan^{-1}w}.
\end{multline}



{\small
\begin{thebibliography}{99}


\bibitem{0012251} L.~Rastelli, A.~Sen, B.~Zwiebach,
\textit{String field theory  around the tachyon vacuum},
hep-th/0012251.



\bibitem{RZ} L.~Rastelli and B.~Zwiebach,
\textit{Tachyon potentials, star products and universality}, JHEP
0109 (2001) 038, hep-th/0006240.

\bibitem{0111129} D.~Gaiotto, L.~Rastelli, A.~Sen and B.~Zwiebach,
\textit{Ghost Structure and Closed Strings in Vacuum String Field
Theory}, hep-th/0111129.

\bibitem{Kostelecky-Potting}
V.A.~Kostelecky and R.~Potting, \textit{Analytical construction of
a nonperturbative vacuum for the open bosonic string},
Phys.Rev.~D63 (2001) 046007, hep-th/0008252.


\bibitem{0102112} L.~Rastelli, A.~Sen, B.~Zwiebach,
\textit{Classical Solutions in String Field Theory Around the
Tachyon Vacuum}, hep-th/0102112.

\bibitem{0105058} L.~Rastelli, A.~Sen and B.~Zwiebach,
\textit{Half-strings, projectors, and multiple D-branes in vacuum
string field theory}, JHEP 0111 (2001) 035, hep-th/0105058.

\bibitem{GT01} D.J.~Gross and W.~Taylor,
\textit{Split string field theory. I}, JHEP 0108, 009 (2001),
hep-th/0105059.


\bibitem{GT02} D.J.~Gross and W.~Taylor,
\textit{Split string field theory. II}, JHEP 0108, 010 (2001),
hep-th/0106036.


\bibitem{0105168} L.~Rastelli, A.~Sen and B.~Zwiebach,
\textit{Boundary CFT construction of D-branes in vacuum string
field theory}, JHEP 0111 (2001) 045, hep-th/0105168.


\bibitem{0105129}
T.~Kawano and K.~Okuyama, \textit{Open string fields as matrices},
JHEP 0106 (2001) 061, hep-th/0105129.


\bibitem{0106010} L.~Rastelli, A.~Sen and B.~Zwiebach,
\textit{Vacuum string field theory}, hep-th/0106010.


\bibitem{0105184}
J.R.~David, \textit{Excitations on wedge states and on the
sliver}, JHEP 0107 (2001) 024, hep-th/0105184.


\bibitem{0107101}
K.~Furuuchi and K.~Okuyama, \textit{Comma vertex and string field
algebra}, JHEP 0109 (2001) 035, hep-th/0107101.

\bibitem{Matsuo} Y.~Matsuo,
\textit{BCFT and sliver state}, Phys.Lett. B513 (2001) 195-199,
hep-th/0105175.
\\
Y.~Matsuo,\textit{Identity projector and D-brane in
string field theory}, Phys.Lett. B514 (2001) 407-412,
hep-th/0106027.
\\
Y.~Matsuo,\textit{Projection operators and D-branes
in purely cubic open string field theory}, Mod.Phys.Lett. A16
(2001) 1811-1822, hep-th/0107007.

\bibitem{0108150} H.~Hata and T.~Kawano,
\textit{Open string states around a classical solution in vacuum
string field theory}, JHEP 0111 (2001) 038, hep-th/0108150.

\bibitem{0110124} I.~Kishimoto,
\textit{Some properties of string field algebra}, JHEP 0112 (2001)
007, hep-th/0110124.

\bibitem{0110136} P.~Mukhopadhyay,
\textit{Oscillator representation of the BCFT construction of
D-branes in vacuum string filed theory}, hep-th/0110136.

\bibitem{0110204} N.~Moeller,
\textit{Some exact results on the matter star-product in the
half-string formalism}, hep-th/0110204.

\bibitem{0111034} H.~Hata and S.~Moriyama,
\textit{Observables as twist anomaly in vacuum string field
theory}, hep-th/0111034.

\bibitem{0111069} G.~Moore and W.~Taylor,
\textit{The singular geometry of the sliver}, hep-th/0111069.

\bibitem{0111087}  K.~Okuyama,
\textit{Siegel Gauge in Vacuum String Field Theory},
hep-th/0111087.

\bibitem{0111092} A.~Hashimoto and N.~Itzhaki,
\textit{Observables of String Field Theory}, hep-th/0111092.

\bibitem{0111153} L.~Rastelli, A.~Sen and B.~Zwiebach,
\textit{A Note on a Proposal for the Tachyon State in Vacuum
String Field Theory}, hep-th/0111153.

\bibitem{0112202}  R.~Rashkov, K.S.~Viswanathan,
\textit{A Note on the Tachyon State in Vacuum String Field Theory}, hep-th/0112202.


\bibitem{0111281} L.~Rastelli, A.~Sen and B.~Zwiebach,
\textit{Star Algebra Spectroscopy}, hep-th/0111281.


\bibitem{0112169} I.~Kishimoto, K.~Ohmori,
\textit{CFT Description of Identity String Field: Toward
Derivation of the VSFT Action}, hep-th/0112169.

\bibitem{0201015} K.~Okuyama,
\textit{Ghost Kinetic Operator of Vacuum String Field Theory}, hep-th/0201015.

\bibitem{0201054} J.~Kluson,
\textit{Some Solutions of Berkovits' Superstring Field Theory}, hep-th/0201054.


\bibitem{0201095} M.~Schnabl,
\textit{Wedge states in string field theory}, hep-th/0201095.

\bibitem{0201136} K.~Okuyama,
\textit{Ratio of Tensions from Vacuum String Field Theory}, hep-th/0201136.

\bibitem{0201149} T. Okuda,
\textit{The Equality of Solutions in Vacuum String Field Theory},
hep-th/0201149.

\bibitem{0201177} H.~Hata, S.~Moriyama, S.~Teraguchi,
\textit{Exact Results on Twist Anomaly}, hep-th/0201177.

\bibitem{0201229}  R.~Rashkov, K.S.~Viswanathan,
\textit{A Proposal for the Vector State in Vacuum String Field Theory}, hep-th/0201229.

\bibitem{0202045} J.~Kluson,
\textit{Exact Solutions of Open Bosonic String Field Theory}, hep-th/0202045.

\bibitem{0202139} M.~Schnabl,
\textit{Anomalous reparametrizations and butterfly states in string field theory}, hep-th/0202139.

\bibitem{0202151} D.~Gaiotto, L.~Rastelli, A.~Sen and B.~Zwiebach,
\textit{Star Algebra Projectors}, hep-th/0202151.

\bibitem{0202176} B.~Feng, Y.-H. He and N. Moeller,
\textit{The Spectrum of the Neumann Matrix with Zero Modes}, hep-th/0202176.

\bibitem{0203089} J.~Kluson,
\textit{Marginal Deformations In the Open Bosonic String Field Theory for N D0-branes
}, hep-th/0203089.



\bibitem{0112214} I.Ya.~Aref'eva, A.A.~Giryavets and P.B.~Medvedev,
\textit{NS Matter Sliver}, hep-th/0112214.

\bibitem{0112231} M.~Marino, R.~Schiappa,
\textit{Towards Vacuum Superstring Field Theory: The Supersliver}, hep-th/0112231.

\bibitem{0201197} I.Ya.~Aref'eva, D.M.~Belov and A.A.~Giryavets,
\textit{Construction of the Vacuum String Field Theory on a Non-BPS Brane}, hep-th/0201197.






\bibitem{0106157} I.~Bars,
\textit{Map of Witten's * to Moyal's *}, Phys.Lett. B517 (2001) 436, hep-th/0106157.

\bibitem{0202030} I.~Bars and Y.~Matsuo,
\textit{Associativity Anomaly in String Field Theory}, hep-th/0202030.


\bibitem{0202087}  M.R.~Douglas, H.~Liu, G.~Moore and B.~Zwiebach,
\textit{Open String Star as a Continuous Moyal Product}, hep-th/0202087.



\bibitem{0201159} D.~Gaiotto, L.~Rastelli, A.~Sen and B.~Zwiebach,
\textit{Patterns in Open String Field Theory Solutions}, hep-th/0201159.




\bibitem{Witten} E.~Witten,
\textit{Noncommutative geometry and string field theory},
Nucl.Phys. B268 (1986) 253;
\\
E.~Witten, \textit{Interacting field theory of open superstrings},
Nucl.Phys.~B276 (1986) 291.



\bibitem{0102085}
K.~Ohmori, \textit{A Review on Tachyon Condensation in Open String
Field Theories}, hep-th/0102085.

\bibitem{0109182} P. De Smet,
\textit{Tachyon Condensation: Calculations in String Field Theory
}, hep-th/0109182.

\bibitem{0111208} I.Ya.~Aref'eva, D.M.~Belov, A.A.~Giryavets, A.S.~Koshelev,
P.B.~Medvedev, \textit{Noncommutative Field Theories and
(Super)String Field Theories}, hep-th/0111208.



\bibitem{GMS} R.~Gopakumar, S.~Minwalla and A.~Strominger,
\textit{Noncommutative Solitons}, JHEP 0005 (2000) 020,
hep-th/0003160.


{\bibitem{0002211} N.~Berkovits,                                                              }
\textit{Super-Poincare Invariant Superstring Field Theory}, Nucl.Phys. B450 (1995) 90,
hep-th/9503099.
\\
N.~Berkovits, A.~Sen, B.~Zwiebach,
\textit{Tachyon Condensation in Superstring Field Theory}, Nucl.Phys. B587 (2000) 147,
hep-th/0002211.



\bibitem{0011117} I.Ya.~Aref'eva, D.M.~Belov, A.S.~Koshelev and P.B.~Medvedev,
\textit{Tachyon Condensation in the Cubic Superstring Field
Theory}, hep-th/0011117.

\bibitem{AMZ1} I.Ya.~Aref'eva,  P.B.~Medvedev and A.P.~Zubarev,
\textit{New representation for string field solves the consistency
problem for open superstring field}, Nucl.Phys.~B341 (1990) 464;
\textit{Background formalism for superstring field theory},
Phys.Lett.~ B240 (1990) 356.
\bibitem{PTY} C.R.~Preitschopf, C.B.~Thorn and S.A.~Yost,
\textit{Superstring Field Theory}, Nucl.Phys.~B337 (1990) 363.




\bibitem{GrJe3} D.~Gross and A.~Jevicki,
\textit{Operator Formulation of Interacting String Field Theory},
(I), (II), Nucl.Phys.~B283 (1987) 1;
Nucl.Phys~B287 (1987) 225.
\\
D.~Gross and A.~Jevicki,
\textit{Operator Formulation of Interacting String Field Theory
(III). NSP superstring}, Nucl.Phys.~B293 (1987) 29.

\bibitem{l'Clare} A.~LeClair, M.E.~Peskin and  C.R.~Preitschopf,
\textit{String field theory on the conformal plane I},
Nucl.Phys.~B317 (1989) 411.
\\
A.~LeClair, M.E.~Peskin and  C.R.~Preitschopf,
\textit{String field theory on the conformal plane II},
Nucl.Phys.~B317 (1989) 464.


\end{thebibliography}
}
\end{document}